\documentclass[aps,prx,reprint,twocolumn,superscriptaddress,floatfix,longbibliography]{revtex4-2}
\usepackage{amsthm}
\usepackage{amsmath,amssymb,color,comment,physics}
\usepackage[makeroom]{cancel}
\usepackage[caption=false]{subfig}
\usepackage{mathrsfs}
\usepackage{graphicx}
\usepackage{float}
\usepackage{subfig}
\usepackage[countmax]{subfloat}
\usepackage[english]{babel}
\usepackage{dsfont}
\usepackage[bookmarks=true,colorlinks,linkcolor=RoyalBlue,urlcolor=NavyBlue,citecolor=RoyalBlue]{hyperref}
\usepackage[dvipsnames]{xcolor}
\usepackage{braket}
\usepackage{mathtools}
\usepackage{bm}
\usepackage{siunitx}

\newcommand{\bv}{\vb{G}}
\newcommand{\rv}{\vb{r}}

\newcommand{\kv}{\vb{k}}

\newcommand{\qv}{\vb{q}}
\newcommand{\Xh}{\hat{X}}
\newcommand{\Xhd}{\hat{X}^\dagger}

\newcommand{\vacbra}{\bra{0}}

\newcommand{\dhd}{\hat{d}^\dagger}
\renewcommand{\dh}{\hat{d}}
\newcommand{\zerov}{\vb{0}}
\newcommand{\Kv}{\vb{K}}
\newcommand{\Qv}{\vb{Q}}
\newcommand{\red}{}

\renewcommand{\dh}{\hat{d}}

\newcommand{\Dfreeh}{\hat{D}^{(0)}}
\newcommand{\Gfreeh}{\hat{G}^{(0)}}

\newcommand{\Dh}{\hat{D}}
\newcommand{\Pih}{\hat{\Pi}}
\newcommand{\area}{\mathcal{A}}

\begin{document}
\title{AC-Stark spectroscopy of interactions between moiré excitons and polarons}

\author{B. Evrard}
\affiliation{Institute for Quantum Electronics, ETH Z\"{u}rich, CH-8093 Z\"{u}rich, Switzerland}
\author{H.S. Adlong}
\affiliation{Institute for Quantum Electronics, ETH Z\"{u}rich, CH-8093 Z\"{u}rich, Switzerland}
\affiliation{Institute for Theoretical Physics, ETH Zürich, Zürich, Switzerland}
\author{A.A. Ghita}
\affiliation{Institute for Quantum Electronics, ETH Z\"{u}rich, CH-8093 Z\"{u}rich, Switzerland}
\author{T. Uto}
\affiliation{Institute for Quantum Electronics, ETH Z\"{u}rich, CH-8093 Z\"{u}rich, Switzerland}
\author{L. Ciorciaro}
\affiliation{Institute for Quantum Electronics, ETH Z\"{u}rich, CH-8093 Z\"{u}rich, Switzerland}
\author{K. Watanabe}
\affiliation{Research Center for Electronic and Optical Materials, NIMS, 1-1 Namiki, Tsukuba 305-0044, Japan}
\author{T. Taniguchi}
\affiliation{Research Center for Electronic and Optical Materials, NIMS, 1-1 Namiki, Tsukuba 305-0044, Japan}
\author{M. Kroner}
\affiliation{Institute for Quantum Electronics, ETH Z\"{u}rich, CH-8093 Z\"{u}rich, Switzerland}
\author{A. {\.I}mamo{\u{g}}lu}
\affiliation{Institute for Quantum Electronics, ETH Z\"{u}rich, CH-8093 Z\"{u}rich, Switzerland}

\begin{abstract}
We use nonlinear pump--probe spectroscopy to study optical excitations in a charge-tunable MoSe$_2$/WS$_2$ moiré heterostructure. An intense red-detuned laser pulse creates a photonic dressing of the material by introducing a large {\sl virtual} population of excitons or exciton--polarons in a deep moiré potential. By measuring the resulting AC-Stark effect with a weak resonant laser pulse, we gain access to the nature and mutual interactions of the elementary optical excitations. At charge neutrality, our measurements reveal that different exciton resonances, associated with confinement of their center-of-mass motion in the moiré potential, have a significant spatial overlap. The resulting short-range interactions manifest themselves as a density-dependent blue shift for same-valley excitons, and bound biexciton states for opposite-valley excitons. The attractive polaron resonance that appears upon injection of electrons into the heterostructure shows a contrasting behavior: here, we observe an electron-density-independent light shift and a clear pump-power-dependent saturation. These features are equivalent to that of an ensemble of independent two-level emitters and indicate a breakdown of the Fermi-polaron picture for optical excitations of electrons subject to a strong moiré potential. Our work establishes an experimental approach to elucidate the elementary optical excitations of semiconductor moiré heterostructures, providing a solid ground for the spectroscopy of correlated electronic and excitonic states in such materials.
\end{abstract}

 \maketitle

\section{Introduction}

Semiconductor moiré materials have emerged as a rich playground for exploration of strongly correlated electrons~\cite{mak2022moiré,andrei2021marvels,Kennes2021review}. \red{In twisted bilayers of transition metal dichalcogenides (TMDs), linear spectroscopy has enabled the observation of a wealth of many-body states, ranging from correlated Mott--Wigner states~\cite{Shimazaki2020,Tang2020,regan2020mott}, through kinetic magnetism~\cite{Ciorciaro2023} to fractional Chern insulators~\cite{Park2023chern,Zeng2023chern}. Indeed, attractive and repulsive exciton-polarons, resulting from the dynamical dressing of excitons by itinerant charges~\cite{Rapaport2000, suris2003correlation, sidler2017fermi,efimkin2017many, glazov2020optical, wang2018colloquium}, provide built-in sensors for both the charge and magnetic order of electrons in TMDs~\cite{Xu2020,smolenski2021signatures}. Going beyond a mere diagnostic purpose, optical excitations of moiré materials have also been proposed as building blocks of correlated bosonic systems \cite{excitonMott,gao2023excitonicMott,Lian2023BFmixture,Park2023DipoleLadders}. Consequently, a complete characterization of moiré exciton and exciton-polaron resonances is essential for correctly interpreting spectroscopic signatures of a broad set of correlated states in semiconductor moiré materials~\cite{lin2023remarkably,wu2017topological,brem2020tunable,naik2022intralayer,Huang2023NonBosonic,Guo2020}.}

Here, we use nonlinear ac-Stark spectroscopy to measure the interactions and possible bound states arising between different elementary optical excitations of a MoSe$_2$/WS$_2$ heterostructure.
Our approach relies on an intense red-detuned pump beam to generate a significant density of virtual excitations, and on a weak broadband probe pulse to monitor the subsequent modification of the optical spectrum. By detuning the pump away from any resonances, we alleviate real-absorption-induced modification of the electronic state and look at the coherent scattering response of the system. 
In TMD monolayers, repulsive Coulomb exchange interactions between itinerant excitations generated by the pump and the probe result in a blue shift when both lasers are co-circularly polarized~\cite{cunningham2019StarkEffect,Slobodeniuk2023,combescot1992semiconductors,schmitt1986collective,zimmermann1990dynamical,haug2009book,uto2024,sarkar2024subwavelengthopticallattice2d}. For cross-circular polarization, the sign of the light shift depends on the pump frequency: This is a manifestation of an underlying Feshbach resonance, occurring when the pump detuning from the exciton resonance equals the biexciton binding energy~\cite{yong2018biexcitonic,sie2016biexciton,hao2017biexciton,cunningham2019resonant,cam2022symmetry}. Our prior measurements also revealed a striking electron density dependence of the light shift which allowed us to determine a dramatic enhancement of interactions between attractive polarons (AP)\,\cite{uto2024}. In this work, we employ ac-Stark spectroscopy on a TMD bilayer featuring a deep moiré potential to reveal three new features: (1)~At charge neutrality where two bright moiré excitons dominate the spectrum, we observe a blue shift of a moiré exciton mode induced by the interaction with another mode. This inter-species ac-Stark effect, with no equivalent in a monolayer TMD, reveals the extent of spatial overlap between the two orthonormal modes. (2)~We identify moiré biexciton Feshbach resonances associated with bound states of both the same and different moiré exciton modes~\cite{Brem2024}. To explain the nature of these biexciton states, we develop a theoretical model for the scattering of excitons in a moiré potential and find good qualitative agreement with the experiment. (3)~We find that the electron density dependence of the ac-Stark shift, as well as the pump-power-dependent saturation, of the AP resonance is qualitatively different from that of a monolayer. Our findings correspond to those of an ensemble of independent two-level emitters, indicating a breakdown of the Fermi-polaron picture for optical excitations of electrons subject to a strong moiré potential.

\begin{figure*}
	\centering
	\includegraphics[width=\linewidth]{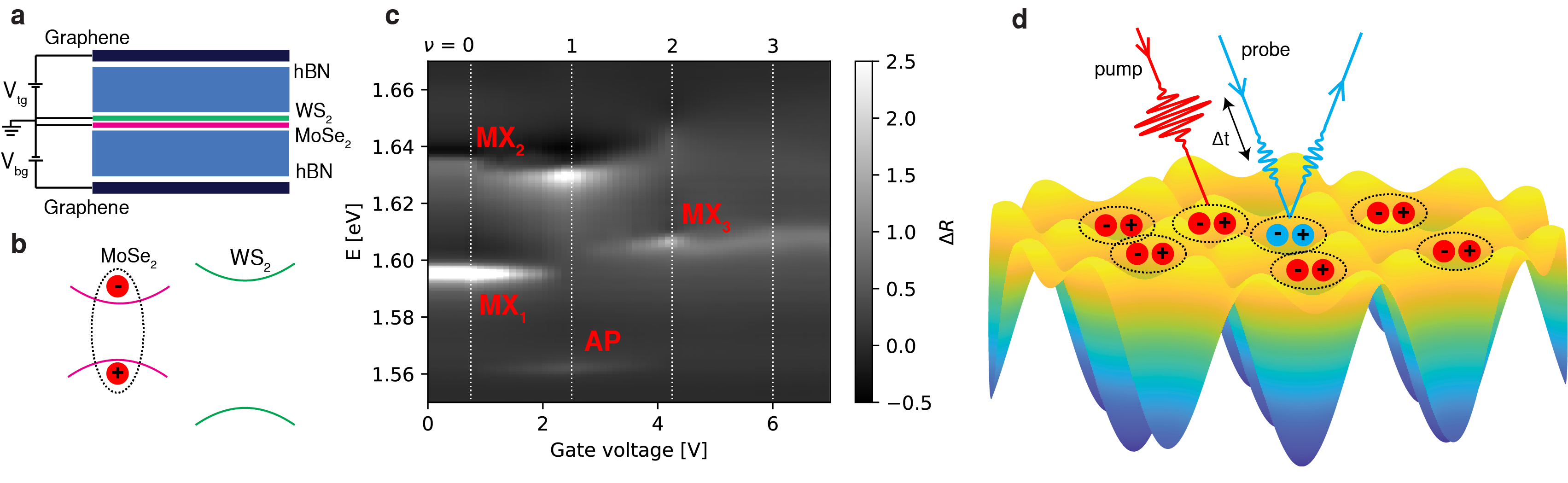}
	\caption{Overview of the sample and experiment. \textbf{a} Schematic of the device. The TMDs MoSe$_2$ and WS$_2$ form the moiré material that we investigate. It is encapsulated in two $\approx 35\,$nm thick hBN flakes,  and dual gated with graphite electrodes to enable an independent control of the chemical potential and electric field. The latter is kept close to zero, such that only MoSe$_2$ is doped at low densities, given the band alignment ({\bf b}). The evolution of the reflection spectrum of MoSe$_2$ as a function of the electronic density ({\bf c}) reveals several moiré exciton and polaron resonances, which we investigate. We rely on a pump--probe scheme ({\bf d}) where an intense red-detuned laser generates a virtual population of moiré excitations. The interaction between this background and a test excitation generated with a probe laser are then measured.}
	\label{Figure1}
\end{figure*}

 Our measurements are carried out in a $\simeq 0^{\circ}$-degree stacked MoSe$_2$/WS$_2$ heterostructure, exhibiting a Type~I band alignment where the lowest (highest) energy moiré conduction (valence) band resides in MoSe$_2$ (Fig.~\ref{Figure1}~{\bf a},~{\bf b}). The electron density dependent reflection contrast exhibits four bright resonances, which we identify as MX$_1$, MX$_2$, AP and MX$_3$ to be consistent with the notation  used in an earlier publication~\cite{Ciorciaro2023} (Fig.~\ref{Figure1}~{\bf c}). A strong pump laser with a finite red detuning from a given resonance generates a large virtual population of the corresponding moiré exciton species/modes that exists only during the $\approx 0.2$\,ps duration of the pump pulse. We estimate a maximum virtual exciton population in MX$_1$ of $\approx 10^{12}$\,cm$^{-2}$ for a pump detuning of $\delta_1\approx 20$\,meV (for more details, see the Appendix). A weak, broadband probe pulse then measures the energy shift of all excitonic species concurrently~(Fig.~\ref{Figure1}~{\bf d}). Throughout this work, we set the pump laser detuning from the excitonic resonances to be much smaller than the exciton binding energy: In this limit, the dominant contribution to the light shift of the resonances originates from exciton--exciton interactions~\cite{cunningham2019StarkEffect,Slobodeniuk2023,combescot1992semiconductors,schmitt1986collective,zimmermann1990dynamical,haug2009book,uto2024}. We determine both intra- and inter-species interaction strengths between the moiré exciton or polaron modes as function of the electron density by measuring the light shift in this small detuning limit.

\section{Interactions between moiré excitons}

We begin by performing non-linear ac-Stark spectroscopy at charge neutrality in order to understand how the moiré potential modifies exciton--exciton interactions and biexcitons. Since the exciton binding energy (Bohr radius) is much larger (smaller) than all other relevant energy (length) scales, the role of the moiré potential is to induce a periodic potential for the center-of-mass motion of the 1s exciton~\footnote{Recent work has shown that the lowest energy moiré exciton resonances can indeed be characterized as 1s excitons~\cite{naik2022intralayer}}. When the moiré potential is weak, we would observe Umklapp resonances, blue shifted from the $k=0$ 1s exciton mode by $\simeq (\hbar k_M)^2/(2m_{\rm ex})$ where $k_M$ is the reciprocal moiré lattice wave vector and $m_{\rm ex}$ is the exciton effective mass. In the opposite limit, we expect resonances corresponding to exciton modes localized around the local minima of the moiré potential. In  charge neutral MoSe$_2$/WS$_2$ heterostructures two bright moiré exciton resonances MX$_1$ and MX$_2$ are observed in the normalized reflection spectrum $\Delta R = (R - R_\mathrm{bg})/R_\mathrm{bg}$, where $R$ and $R_\mathrm{bg}$ denote the reflection spectrum from the sample and background, respectively (see Fig.~\ref{Figure1}\textbf{c}). The splitting between these two peaks of $\approx 41\,$meV is significantly larger than the expected Umklapp splitting $(\hbar k_M)^2/(2m_{\rm ex})\approx 27\,$meV for this system, which suggests that the moiré potential is strong enough to significantly alter the exciton spatial wave-function within the moiré unit cell.

\red{To reveal the elementary features of moiré excitons, such as their spatial extent and overlap, we investigate their mutual interactions using ac-Stark spectroscopy. By measuring the light shift for co- and cross-circularly polarized pump and probe beams, we probe interactions between moiré excitons in the same (co-circular) or opposite valley (cross-circular).}

\subsection{Co-circularly polarized light}

\begin{figure*}
	\centering
	\includegraphics[width=\linewidth]{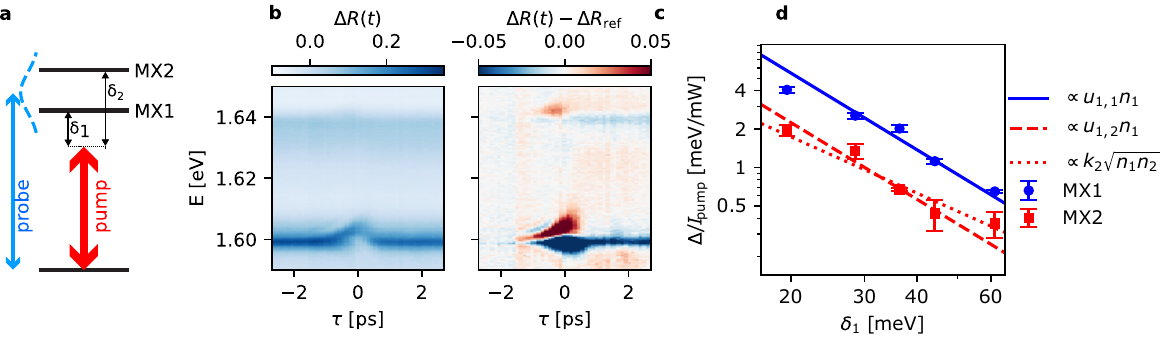}
	\caption{\textbf{Light shift of moiré excitons for co-circularly polarized pump and probe lasers.} (\textbf{a})~Schematic of the moiré exciton energy level at charge neutrality. (\textbf{b})~Reflection spectrum as a function of the pump--probe delay, showing a clear blue shift of the lowest and brightest resonance MX$_1$ at zero time delay. The differential spectrum (\textbf{c}), obtained by subtracting from (\textbf{b}) a reference spectrum (acquired at $\tau<-2.5\,$ps), reveals a blue shift of the upper and darker resonance MX$_2$ as well. (\textbf{d})~The dependence of the blue shift of MX$_1$ (blue circles) on the pump detuning scales as $1/\delta_{1}^2\propto n_1$ (blue line), and stems from MX$_1$--MX$_1$ interactions. The shift of MX$_2$ (red squares), is well fitted either by a $\propto1/\delta_{1}^2$ dependence (dashed red line) or by $\sqrt{n_1n_2}\propto1/(\delta_{1}\delta_{2})$ (dotted red line). While we are not able to precisely deconvolve the role of these two potential contributions, both imply significant MX$_1$--MX$_2$ interaction and hence spatial overlap between these two bright moiré excitons.}
	\label{Figure2}
\end{figure*}

Beginning with the scenario of co-circular polarization, we consider the evolution of $\Delta R$ as a function of the delay $\tau=t_{\rm probe}-t_{\rm pump}$, between the pump and probe pulse, as shown in Fig.~\ref{Figure2}~{\bf b} (here the pump pulse is red-detuned from the MX$_1$ resonance by $\delta_{1} = 20$\,meV).
The dominant coupling mechanisms between two same-valley excitons are electron and hole exchange interactions~\cite{ciuti1998interaction,shahnazaryan2017excitoninteraction}, which lead to short-range repulsion. Correspondingly, we observe a clear blue shift of the brightest exciton MX$_1$ for $|\tau|\lessapprox0.2\,$ps, that is, when the two pulses overlap in time (for more detail on the experimental setup see the Appendix). To better assess the pump-induced modifications of weaker resonances, we use the differential reflection spectrum $\Delta R(t)-\Delta R_{\rm ref}$, where $\Delta R_{\rm ref}$ is a reference spectrum obtained when the probe pulse hits the sample significantly before the pump ($\tau\lessapprox-2.5\,$ps): Figure~\ref{Figure2}~{\bf c} shows that a smaller blue shift of MX$_2$ is discernible in $\Delta R(t)-\Delta R_{\rm ref}$.  In a mean-field picture (detailed in App.\,\ref{App: Same valley interactions}), the light shift $\Delta_i$ of MX$_i$ ($i=1,2$) can be expressed as
\begin{subequations} \label{Eq:MFShifts}
    \begin{align}
    \Delta_1 = 2u_{1,1} n_1+2 u_{1,2} n_2+4 k_1 \sqrt{n_1n_2}\,,\\
    \Delta_2 = 2 u_{2,2} n_2+2u_{1,2}n_1+ 4 k_2\sqrt{n_1n_2}\,,
\end{align}
\end{subequations}
where $n_i$ is the density of MX$_i$ excitons and we have introduced four different interaction terms: $u_{i,j}$ correspond to exciton--exciton scattering conserving the moiré miniband populations, i.e.\ the processes MX$_i$+MX$_j\leftrightharpoons$ MX$_i$+MX$_j$, while $k_i$ correspond to processes where an exciton is scattered into a different moiré miniband, i.e.\ MX$_1$+MX$_2\leftrightharpoons$ MX$_i$+MX$_i$. The density scales as $n_i\propto1/\delta_i^2$ with the detuning $\delta_i$ of the pump from the MX$_i$ resonance. Therefore, by tuning the pump laser frequency, one can change the density imbalance between the two excitons, and in principle deconvolve the contribution of each scattering processes to the light shift. In practice, we observe signatures of an incoherent response for a blue-detuned pump, so we focus exclusively on red detunings. We consequently always have $\delta_1<\delta_2$ and thus $n_1>n_2$ (see Fig.~\ref{Figure2}~{\bf a}). This imbalance is further amplified by the oscillator strength difference between MX$_1$ and MX$_2$.

Figure~\ref{Figure2}~{\bf d}  shows the light shifts $\Delta_{1,2}$ as a function of $\delta_1$ in a range where $n_1\gtrsim 10\,n_2$. As a result, the light shift of MX$_1$ is dominated by MX$_1$--MX$_1$ interactions. From a fit we obtain the interaction strength $u_{1,1}$, which we find to be larger by a factor $u_{1,1}/u_{\rm ex}\approx 1.6\pm0.2$ than the moiré-free exciton--exciton interaction strength $u_{\rm ex}$, measured on a monolayer MoSe$_2$ region of the same device. We tentatively attribute this enhancement  to the reduced spatial extent of the MX$_1$  center-of-mass wavefunction within the moiré unit cell, which in turn increases the overlap between MX$_1$ excitons for a given average density.

Remarkably and despite of the large imbalance $n_2\ll n_1$, we observe a substantial light shift of MX$_2$, strongly increasing as the pump wavelength approaches the MX$_1$ resonance. This shift $\Delta_2$ is not well reproduced by a $1/\delta_2^2$ dependence, and attests to substantial inter-species interactions. A fit to our data does not enable us to precisely disentangle the respective contributions of the MX$_2$+MX$_1\leftrightharpoons$ MX$_2$+MX$_1$ and MX$_2$+MX$_1\leftrightharpoons$ 2MX$_1$ processes. Nevertheless, we estimate that $0.3\lesssim u_{1,2}/u_{1,1}\lesssim 0.8$ while we are not able to reliably estimate the $k_i$ parameters (see App.\,\ref{App: Fit} for more details on the fitting procedure).

These observations suggest a significant spatial overlap of MX$_1$ and MX$_2$ excitons and invalidates a simplistic picture of moiré exciton modes that are tightly confined around different high-symmetry points of the moiré potential. Moreover, our experiments are in reasonable qualitative agreement with the findings of moiré exciton wavefunctions obtained for the same structure using the continuum model \cite{polovnikov2022coulomb}. In particular, we find that the latter predicts $u_{1,1}/u_{\rm ex} = 2.2$ and  $u_{1,2}/u_{1,1}= 0.6$ (see Appendix~\ref{App:TheoryModel} for details).

We point out that Eq.~\eqref{Eq:MFShifts} can be understood as arising from first order perturbation theory. This approach is valid in the limit of  sufficiently small pump power/large pump detuning. We find an empirical confirmation that we are indeed working in that regime by observing a linear dependence of the light shifts $\Delta_i$ with the pump intensity (App.\,\ref{App: DataAnalysis}). Nevertheless, we can envision interesting higher order effects arising from the mixing of the excitonic states. In particular, mixing of optically bright and dark excitons, could be observed as the emergence of new resonances in the reflection spectrum.

\begin{figure*}
    \centering
    \includegraphics[width=\linewidth]{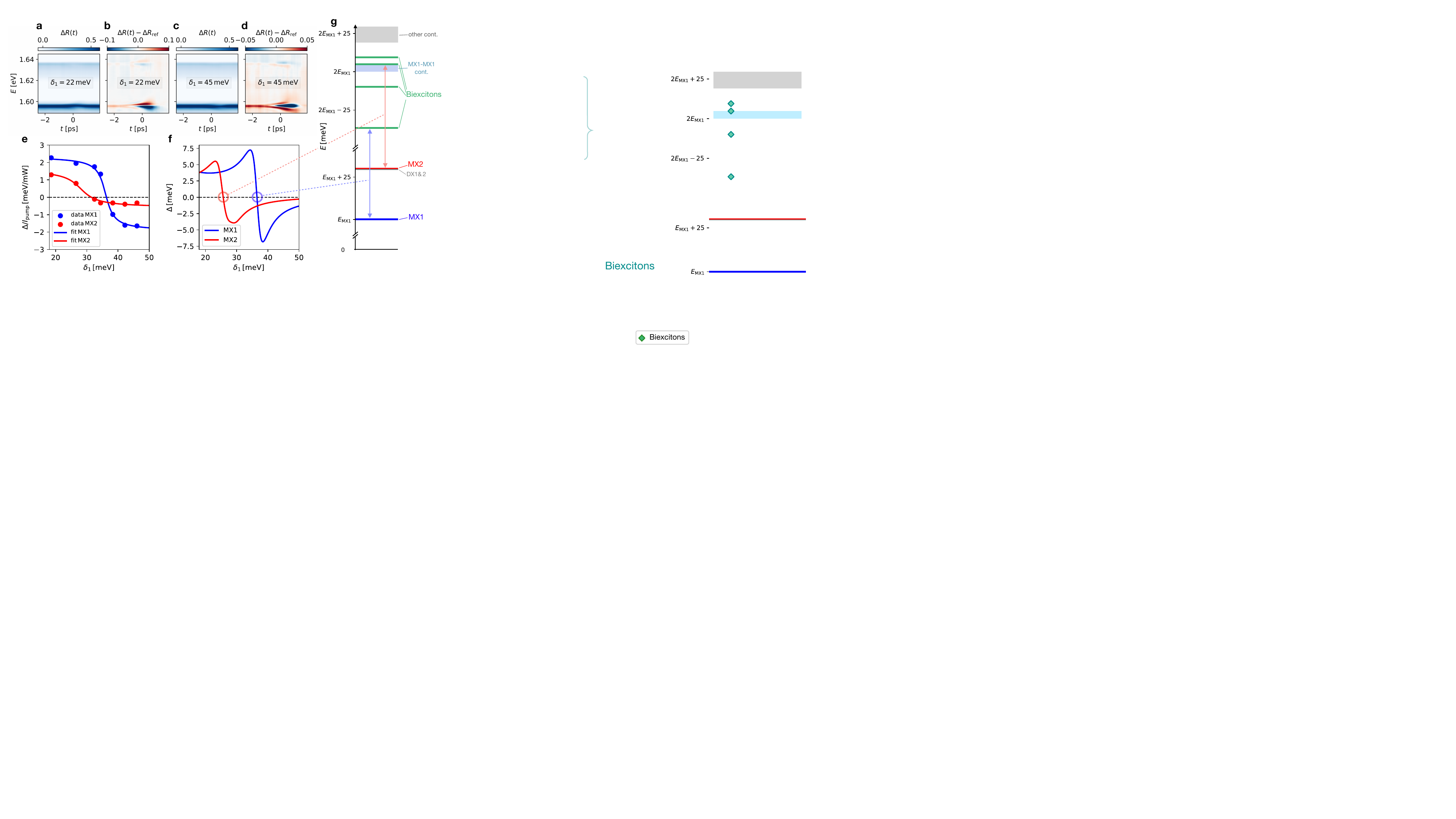}
    \caption{\textbf{Coupling to biexcitonic states.} (\textbf{a}) Reflection spectrum as a function of the pump--probe delay when the pump is red-detuned by $22\,$meV from MX$_1$ and the pump and probe laser are cross-circularly polarized. (\textbf{b}) The differential spectrum, obtained by subtracting from (\textbf{a}) a reference spectrum (acquired at $t<-2.5\,$ps), reveals a blue shift of both resonances MX$_1$ and MX$_2$. (\textbf{c},~\textbf{d}) A similar measurement carried out for a detuning of $45\,$meV from MX$_1$ shows instead a red shift of both resonances. (\textbf{e}) The dependence of the light shifts of MX$_1$ (blue dots) and MX$_2$ (red dots) displays a sign change for a pump detuning $\delta_{1}\approx 36\,$meV and  $\delta_1\approx 28\,$meV, respectively. (\textbf{f}) The corresponding theory simulation for the light-shifts of MX$_1$ (blue) and MX$_2$ (red) based on DFT moiré parameters. (\textbf{g}) The calculated spectrum which is shown in three different energy sectors with zero (defined as zero energy), one and two excitons. In the one-exciton sector there are the optically bright MX$_1$ and MX$_2$ excitons. In the two-exciton sector (focusing on zero center-of-mass quasi-momentum) there exists a band of MX$_1$--MX$_1$ excitons (light blue) as well as other bands (gray), and bound biexcitons (green lines). We find that the sign change in the light shift of MX$_1$ and MX$_2$ in \textbf{(f)} can be attributed to the blue and red arrows (indicating exciton--biexciton transitions), respectively.
    }\label{fig2SI}
\end{figure*}

\subsection{Cross-circularly polarized light}
Since electron and hole exchange interactions are suppressed for excitons generated in opposite valleys, one may naively assume that the cross-circularly polarized scenario should yield a significantly smaller light shift. However, the bare interaction of opposite valley excitons is attractive, and supports a bound biexciton state, which has significant implications.

In order to illustrate how the biexciton state affects the exciton--exciton interactions, we briefly review the simpler scenario in which there is no moiré potential (such as in the case of monolayer TMDs). In this case it is known that the presence of the biexciton resonance leads to an additional contribution to the light shift, scaling linearly with the inverse of the two-photon detuning $\delta_{\rm b}^{-1}=(-E_{\rm b}+\delta_{\rm ex})^{-1}$, where $\delta_{\rm ex}$ is the detuning of the pump from the exciton resonance, $E_{\rm b}$ is the biexciton binding energy and the probe is assumed resonant with the exciton.
A hallmark of the biexciton is the ac-Stark effect, where the light shift changes sign for detunings in the vicinity of $E_{\rm b}$~\cite{yong2018biexcitonic,sie2016biexciton,hao2017biexciton,cunningham2019resonant,cam2022symmetry}.
The two-photon resonance condition $\delta_{\rm b} =0$ can be considered as a biexciton Feshbach resonance where the effective interactions between the pump- and probe-generated excitons changes from being attractive  ($\delta_{\rm b} > 0$) to repulsive ($\delta_{\rm b} < 0$).

We now extend these concepts to study the nature of biexcitons in moiré materials by measuring the ac-Stark effect of MX$_1$ and MX$_2$ excitons under cross-circularly-polarized pump--probe lasers as a function of $\delta_1$.
We first observe, that the sign of the ac-Stark shift changes from a blue shift
at small detunings ($\delta_1\approx22$\,meV, Fig.~\ref{fig2SI}~{\bf a} and {\bf b}) to a red shift at large detunings ($\delta_1\approx45$\,meV, Fig.~\ref{fig2SI}~{\bf c} and {\bf d}) for both moiré excitons.
The full detuning dependence in Fig.~\ref{fig2SI}~{\bf e} reveals that the sign change
occurs at different detunings for MX$_1$ and MX$_2$ and is associated with two
different \textit{zero quasi-momentum} biexcitons. By contrast, a monolayer
hosts only a single \textit{zero momentum} exciton. From a heuristic fit with
the function $\Delta_i = a \tan^{-1}{\left(\frac{\delta_1-E_{\mathrm{b},i}}{b}\right)}+c$  (with $i=1,2$ and $a,b,c$ fitting parameters), we extract the binding energies $E_{\rm b,1}$ and $E_{\rm b,2}$ of the biexcitonic states determining the ac-Stark shifts of MX$_1$ and MX$_2$, respectively.
From the MX$_1$ light shift, we find $E_{\rm b,1}\approx36\,$meV, which suggests that the ground state moiré biexciton (MX$_1$--MX$_1$) is more strongly bound here compared to a monolayer MoSe$_2$, where $E_{\rm b}$ was measured to be $\approx20\,$meV~\cite{yong2018biexcitonic,hao2017biexciton} or $\approx29$\,meV \cite{uto2024,tan2022bose} (these variations could stem from different device architectures and dielectric environments).
The light shift of MX$_2$ is related to a second biexciton state, with a mixed MX$_1$/MX$_2$ character and a binding energy $E_{\rm b,2} \approx 28\,$meV measured with respect to an unbound MX$_1$ and MX$_2$ exciton. Importantly, $E_{\rm b,2}$ is smaller than the energy splitting between MX$_1$ and MX$_2$ ($\approx40$\,meV), and this second biexciton is higher in energy than two unbound MX$_1$ excitons. This should be contrasted with the monolayer scenario, where the bound state by definition has lower energy than two unbound zero-momentum excitons.

We further investigate the biexciton states of moiré excitons by modeling their scattering interactions in a periodic potential (see details in Appendix~\ref{App:TheoryModel}). Our model incorporates short-range interactions calibrated to match the biexciton binding energy: we use the lower estimate $E_{\rm b}\approx20\,$meV~\cite{yong2018biexcitonic,hao2017biexciton} to account for screening effects from the proximal WS$_2$. Using parameters derived from Density Functional Theory (Appendix~\ref{App:TheoryModel}), we identify the bright excitonic resonances observed in experiment, which we show in Fig.~\ref{fig2SI}\textbf{(g)}. Here we also show the bound biexcitons, which emerge due to spatial overlap of exciton wavefunctions within the moiré lattice.

The biexciton spectrum reveals that higher-energy biexcitons form in gaps within the two-exciton continuum, consistent with experimental observations. Specifically, the bound MX$_1$--MX$_2$ biexciton is enabled by significant spatial overlap between the exciton wavefunctions, with MX$_2$ being notably more delocalized. The pump laser energies for exciting MX$_1$--MX$_1$ and MX$_1$--MX$_2$ biexcitons are highlighted in Fig.~\ref{fig2SI}\textbf{(g)}.

To connect theory with experiment, we calculate the exciton energy shifts induced by the pump laser (see Appendix~\ref{App:TheoryModel}2--3). Here we estimate the light-matter coupling strength such that the induced exciton density is on the order of $10^{11}$ cm$^{-2}$ (see Appendix~\ref{App:TheoryModel}3). The shifts are shown in Fig.~\ref{fig2SI}\textbf{(f)} and agree qualitatively with measurements. The model suggests that only two biexcitons significantly influence the MX$_1$ and MX$_2$ shifts for the measured experimental detunings; the other biexcitons have negligible overlap with these two excitonic resonances. These results highlight how the emergence of inter-species biexcitons in moiré show clear and direct signatures in ac-Stark spectroscopy.

\section{Attractive polaron light shift}
\begin{figure}
	\centering
	\includegraphics[width=\linewidth]{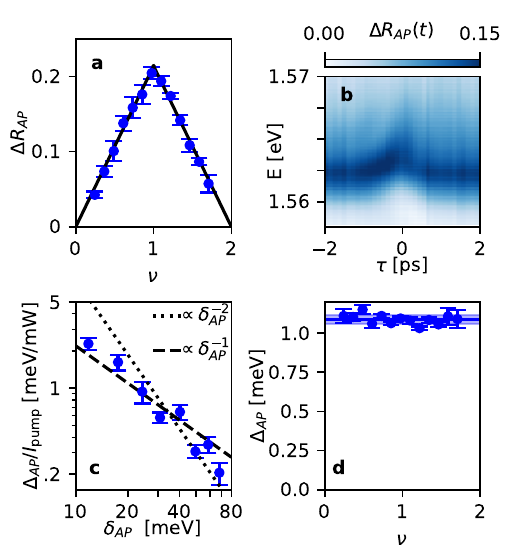}
	\caption{\textbf{Light shift of moiré polarons.} (\textbf{a}) 
 The peak reflection contrast as a function of gate voltage, showing that the oscillator strength of the AP linearly increases until $\nu=1$ and then decreases linearly until $\nu=2$. (\textbf{b}) Reflection spectrum $\Delta R_\mathrm{AP}(\tau)$ of the AP at $\nu =1$ as a function of the pump--probe delay $\tau$. (\textbf{c}) The observed AP blue shift is not driven by interactions as shown by its dependence on the pump detuning. (\textbf{d}) The measured light shift  is remarkably constant as a function of $\nu$, up to a small dispersion $\sim3\%$ (shown as a blue stripe) compatible with statistical fluctuations (shown as error bar, obtained for each point by repeating the measurement four times). The combination of linear $1/\delta_\mathrm{AP}$ dependence of the light shift and its independence of $\nu$ demonstrate the lack of interactions between APs or trions localized on different moiré sites.}
	\label{Figure4}
\end{figure}
When the MoSe$_2$/WS$_2$ heterostructure is electron doped, we observe an AP resonance that is red-shifted from the MX$_1$ by $\approx35\,$meV. The corresponding trion binding energy is about $50 \%$ larger than that observed in monolayer MoSe$_2$, suggesting that the electron Wannier orbitals, as well as the optically generated trions, are strongly localized around the minimum of the moiré potential. Furthermore, the AP oscillator strength follows the number of singly occupied moiré sites, increasing linearly until $\nu=1$ and then decreasing linearly until $\nu=2$ (Fig.~\ref{Figure4} {\bf a}). This observation shows that the AP formation is hindered at doubly occupied sites due to the existence of local electron singlets that Pauli-block the formation of (localized) trions. However, linear spectroscopy cannot be used to identify features that distinguish APs in deep moiré potentials from their counterparts in weak moiré potentials or in monolayer TMDs.

Figure~\ref{Figure4}~{\bf b} shows the pump--probe measurement as a function of $\tau$ for $\delta_\mathrm{AP} = 12$\,meV, clearly showing a blue shift of the AP resonance with negligible alteration of the resonance for $\tau \ge 0.2$\,ps. While this measurement is reminiscent of the result for MX$_1$, Fig.~\ref{Figure4}~{\bf c} shows a striking difference in the $\delta_\mathrm{AP}$ dependence: the AP light shift is better fit using a $1/\delta_\mathrm{AP}$ dependence, typical of the ac-Stark shift observed for an ensemble of non-interacting two-level emitters. We tentatively explain this observation by arguing that the AP resonance can be considered as stemming primarily from a collective excitation of trions localized at the M-M sites of the moiré lattice. In the absence of inter-site hopping, the excitation of a trion at a given site cannot depend on the existence of a trion on any other site, ensuring that moiré trions behave as non-interacting excitations. The only contribution to the light shift in this limit will come from the ac-Stark shift of each site independently, whose magnitude scales as $1/\delta_\mathrm{AP}$. Small but non-zero hybridization of the collective trion excitation with the bare exciton could give rise to a finite interaction strength, a finite inter-site hopping due to long-range electron--hole exchange and a deviation form the pure $1/\delta_\mathrm{AP}$ contribution to the light shift. We note that recent experiments on the same moiré structure yielded magnetization signatures consistent with unexpectedly weak inter-site hopping~\cite{Ciorciaro2023}.

We find the first confirmation of this explanation when we measure the electron density dependence of the light shift: Figure~\ref{Figure4}~{\bf d} shows that varying $\nu$ from $0$ to $2$ results in negligible variation in the magnitude of the light shift, despite large variation in the pump-induced virtual AP population. This behavior contrasts with the strong electron density dependence of the AP light shift previously observed in a monolayer MoSe$_2$ \cite{uto2024}.

\begin{figure}
	\centering
	\includegraphics[width=\linewidth]{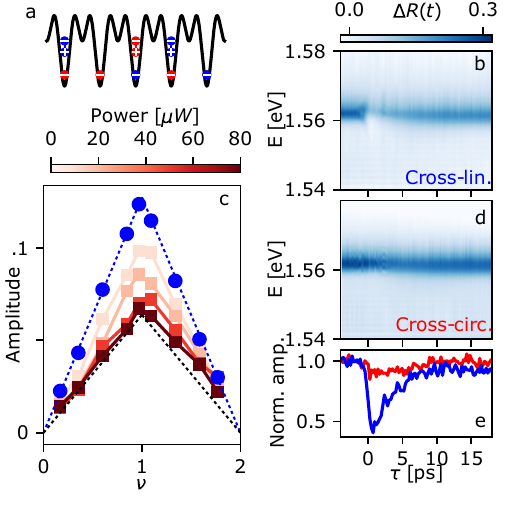}
	\caption{\textbf{Resonant excitation of moiré polarons.} When the pump and probe lasers are cross-linearly polarized, they drive the AP transition at all moiré sites (see the sketch in {\bf a}). We observe a blue shift and a reduction of the AP reflection amplitude ({\bf b}). The latter is plotted in ({\bf c}) as a function of the electronic filling factor $\nu$ of the moiré lattice. The blue dots are obtained without pump and the red squares for various pump intensities. For a sufficiently large intensity, all sites end up in a statistical mixture of a single electron and a moiré trion and consequently the amplitude of the AP resonance is divided by a factor two (dashed black line). After the pump pulse is gone, the AP resonance recovers on a timescale of $\approx 5\,$ps ({\bf e} blue line). In contrast, using cross-circularly polarized lasers, the pump and the probe pulse are driving the AP transition on different moiré sites, and consequently we do not observe any effect of the pump ({\bf d},{\bf e} red line).}
	\label{Figure5}
\end{figure}

An additional confirmation of our description of the moiré AP as a collective excitation of non-interacting localized trions is provided by complementary measurements where we study the saturation behaviour of the AP resonance under resonant pump-laser excitation. Figure~\ref{Figure5}~{\bf a}  depicts a cartoon of the heterostructure where resonant excitation leads to localized trion occupation, which in turn Pauli-blocks further excitation. To investigate the associated saturation of the AP resonance, we measure $\Delta R$ as a function of the electron density in cross-linear configuration for various pump laser intensities.
Figure~~\ref{Figure5}~{\bf b} shows that for pump laser power $P_p = 40\,\mu$W, the AP resonance blue shifts and weakens for short pump--probe time delays. Figure~\ref{Figure5}~{\bf c} in turn, shows that increasing $P_p$ leads to saturation of $\Delta R$  for all electronic densities in the range $0 \le \nu \le 2$. This saturation behavior, characteristic of an ensemble of non-interacting two-level emitters, can be explained by arguing that each occupied moiré site is driven into a balanced mixture of its ground (single electron) and excited (trion) states for $P_p \ge 50 \mu$W. In stark contrast, saturation of composite-boson excitations, such as itinerant excitons or APs, will lead to a gradual loss of oscillator strength, accompanied by a shift of the resonance energy. Figure~\ref{Figure5}~{\bf e} shows that the saturation under cross-linear excitation with $P_p = 40 \mu$W is near-complete for pump--probe time delay $\tau\lesssim\tau_\mathrm{pump}$, where the probe $\Delta R$ is reduced by  a factor $\simeq 2$, as compared to its reference value obtained without pump laser. We remark that the relaxation time of $\approx 5$\,ps of the resonantly excited APs observed in Fig.~\ref{Figure5}~{\bf e} is much longer than the  pump pulse duration $\tau_\mathrm{pump} \simeq 0.1$\,ps. This long-time dynamics is most likely due to the generation of a (real) trion population through absorption of resonant photons.

The aforementioned cross-linearly polarized measurements ensure that both the pump and probe lasers drive trion formation on all moiré sites, regardless of the electron spin. We also performed a test experiment using a cross-circular configuration. In that case, the pump and the probe are addressing singly-occupied sites with electrons in opposite valley. Consequently, the probe reflection is unaffected by the pump (Figure~\ref{Figure5}~{\bf d},{\bf e}). This results stands in contrast to prior measurements on monolayer MoSe$_2$, where in cross-circular configuration we observed a red-shift of the AP resonance, which we tentatively attributed to phase space filling induced by the virtual AP population generated by the pump pulse\,\cite{uto2024}.

We conclude that the pump-induced light shift and bleaching of the AP transition demonstrates that the AP resonance can be described as an ensemble of non-interacting two-level emitters and as such, is inconsistent with the exciton-polaron model that successfully describes the optical spectrum of monolayer TMDs~\cite{sidler2017fermi,efimkin2017many,glazov2020optical} as well as APs in weak moiré potentials~\cite{Kiper2024}.
For electron densities satisfying moiré filling factor $\nu > 1$, another resonance emerges in the spectrum which we term MX$_3$ (Fig.\,\ref{Figure1}\,\textbf{c}). The ac-Stark shift of MX$_3$ is similar to that of monolayer AP or exciton resonances, which allows us to tentatively identify it as excitons or APs generated at doubly occupied moiré sites. These measurements are detailed in the Appendix\,\ref{App: MX3} (see Fig.\,\ref{FigureMX3}): briefly, MX$_3$ displays a density-dependent ac-Stark shift, contrasting with that of the AP discussed above, which we attribute to repulsive interactions. Interestingly, the light shift is smaller for integer fillings $\nu=2$ and $\nu=3$, where the oscillator strength is larger and where the electrons form incompressible states. This counter-intuitive behavior is qualitatively understood from the enhancement of the interaction when the electron system is compressible and efficiently mediates interaction between exciton-polarons\,\cite{uto2024,tan2020interacting}.

\section{Discussion}
Our findings shed new light on the nature of moiré excitons and polarons,  which remains a topic of active research~\cite{huang2022excitons,Du2023review,zhang2018moiré,jin2019observation,liu2021signatures,Tang2020,wu2022evidence,Tran2019,alexeev2019resonantly,Seyler2019,lin2023remarkably,wu2017topological,brem2020tunable,naik2022intralayer,Huang2023NonBosonic,Guo2020,Brem2024}. In particular, we show how non-linear spectroscopy unveils the itinerant or localized character of moiré optical excitations. Our experiments allow us to assess the extent of spatial overlap between different moiré excitons or attractive polarons. It is somewhat remarkable that this information is accessible to far-field optics given that the moiré length scale is about two orders of magnitude below the optical resolution. The insight we obtain is crucial for the interpretation of experiments aimed at optical sensing of correlated electronic states \cite{Xu2020,smolenski2021signatures,Ciorciaro2023,Park2023chern,Shimazaki2020,Tang2020,regan2020mott}. While the heterostructure we studied exhibits a deep triangular moiré potential, favoring topologically trivial correlated electronic insulators and localized AP,  we envision that applying ac-Stark spectroscopy to twisted homobilayers exhibiting Chern bands could reveal features not accessible to linear spectroscopy. For example, we expect the AP ac-Stark shift to change qualitatively as the system is tuned from a fractional Chern insulator quantum fluid to a Mott--Wigner state with strongly localized charges, using an applied displacement field~\cite{Park2023chern,Zeng2023chern}. We also expect the ac-Stark shift of the AP resonance to reveal signatures of the quasi-gap to charged excitations of a composite Fermi liquid at half-filling of a Chern band~\cite{Anderson2024}.

The experiments detailed in Fig.~\ref{Figure5} present a realization of a nonequilibrium Bose--Fermi mixture consisting of electrons in a flat moiré band and optically injected excitons~\cite{excitonMott,gao2023excitonicMott,Lian2023BFmixture,Park2023DipoleLadders}. The choice of resonant excitation of the AP transition forces the mixture into a state that can be described as a high-density moiré trion gas. Using a circularly polarized resonant pump laser and an external magnetic field to valley polarize electrons, it may be possible to create a trion at each moiré site, thereby realizing a solid-state analog of the Dicke model\,\cite{Yu2017}.

The dressing of a quantum material with virtual optical excitations could be a promising route to engineer new phases of matter. The idea of using an intense laser pulse to modify  material properties has already been demonstrated\,\cite{Fausti2011}, but it often suffers from incoherent pumping and heating effects when an electronic polarization mode of the system is resonantly driven. Similar, albeit much less severe, problems in driven atomic systems can be alleviated using Rydberg dressing, where an off-resonant laser effects coherent hybridization of a  ground (or long-lived low-energy) state and a Rydberg state: The atoms remain mostly in the ground state and spontaneous emission is strongly suppressed, but virtual excitations to the Rydberg state still ensure long-range interactions. Similarly, it was proposed that virtual excitons in bulk semiconductors could mediate ferromagnetic interactions between  free electrons in the bulk\,\cite{Piermarocchi2002} or trapped in quantum dots\,\cite{FernndezRossier2004}. Our experiments demonstrate that an intense pump laser can realize an efficient excitonic dressing of a moiré system, while keeping light absorption negligible thanks to a large detuning from the resonances. This approach could provide a new path toward light-induced magnetism in van der Waals heterostructures \cite{Wang2022}.

\section{Acknowledgement}
We thank A. Christianen, M. Glazov, M. Hafezi, A. Müller, A. G. Salvador, R. Schmidt and A. Srivastava for inspiring discussions. This work was supported by the Swiss National Science Foundation (SNSF) under Grant Number 200020$\_$207520. B.\,E. acknowledges funding from an ETH postdoc fellowship. H.\,S.\,A. acknowledges support from the Swiss Government Excellence Scholarship.
T.\,U.~acknowledges support from the Funai Overseas Scholarship.
\\

The data are available at the ETH Research Collection \cite{data_doi}.

\newpage
\section{Appendix}
\appendix

\section{Experimental setup and sample fabrication}

The sample was measured in a dry cryostat (Attodry800, Attocube) at cryogenic temperaures of $\approx 5$K with free-space optical access and equipped with nanopositioners allowing displacement along the three axes. For the pump and probe, we used a mode-locked Ti:sapphire laser (Tsunami, Spectra-Physics), with a repetition rate of 76\,MHz and pulse duration $\approx 100$\,fs. The pulse is split along two paths for the pump and the probe. The bandwidth of the pump is reduced using a pulse shaper and its power is controlled using a motorized optical attenuator. We achieve a larger spectral width for the probe using a nonlinear fiber (femtowhite 800, NKT Photonics) which produces a quasi-continuum around the investigated resonances. The length of the probe optical path can be varied using a retroreflector on a motorized translation stage, enabling a fine tuning of the time delay between the two pulses. Both beams were focused on a diffraction-limited spot on the sample using an apochromatic microscope objective with NA $= 0.8$ (LT-APO/VISIR/0.82, Attocube). The typical probe power is on the order of a few microwatts. The reflected light spectra were recorded using a Peltier-cooled CCD camera.

For the sample fabrication, few-layer graphite, $\approx 35$\,nm hBN, monolayer MoSe$_2$ and WS$_2$ were mechanically exfoliated. The layers were assembled using the dry-transfer technique with a poly(bisphenol A carbonate) film on a polydimethylsiloxane (PDMS) stamp and deposited on a 285\,nm Si/SiO$_2$ substrate \cite{zomer2014fast}. The crystal alignement of the TMDs was determined prior to the stacking measuring the generation of second-harmonic light as a function of the polarization of an incoming infrared laser pulse. They were then stacked with a negligible twist. The graphene top and bottom gates and TMDs were contacted using gold electrodes deposited using optical lithography and electron beam deposition.

\section{Data analysis}\label{App: DataAnalysis}
\begin{figure}
    \centering
    \includegraphics[width=\linewidth]{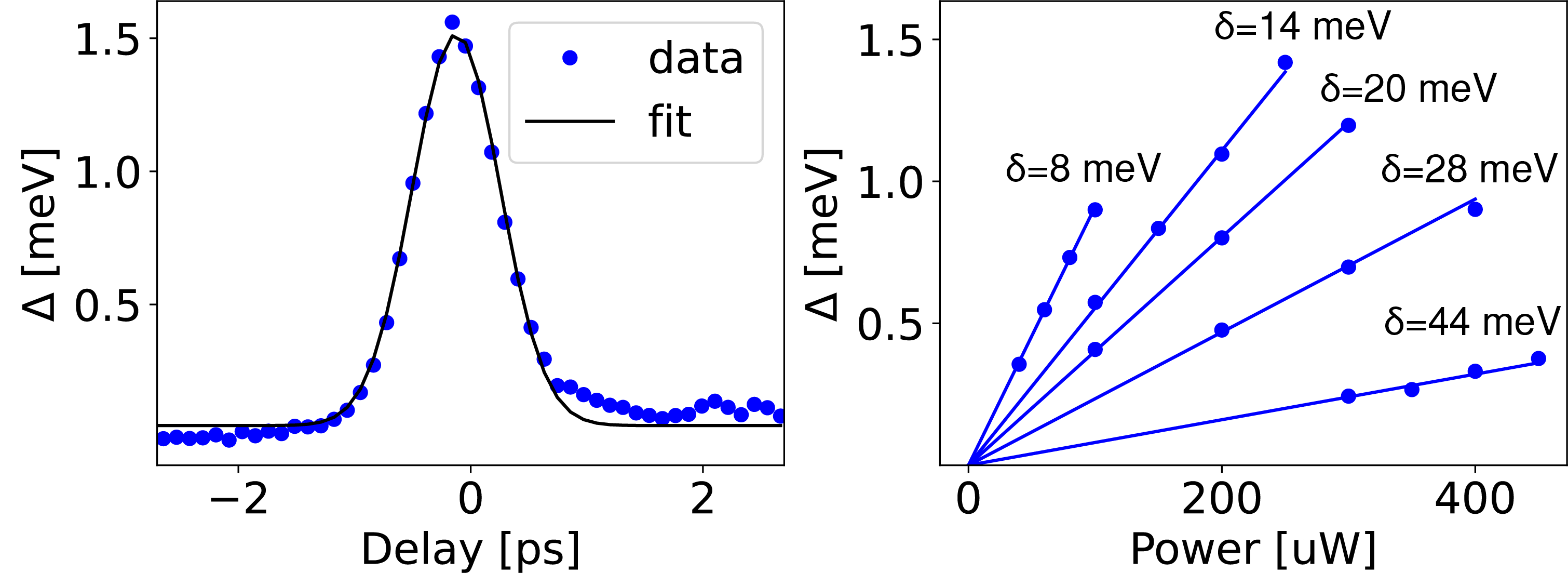}
    \caption{\textbf{Data analysis.} (\textbf{a}) Light shift as a function of the pump--probe delay for MX$_1$ for a pump power of \SI{250}{\micro\watt} and detuning of 11\,meV, showing a clear blue shift at zero time delay. The fit corresponds to a Gaussian envelope. (\textbf{b}) The dependence of the light shift of MX$_1$ on pump power at different pump detunings $\delta_1$ exhibits a linear dependence.}
    \label{fig4SI}
\end{figure}

In order to extract the light shift amplitude, we first fit the reflection spectrum for various time delays and extract the resonance position(s). The result is shown in Fig.\,\ref{fig4SI}\,{\bf a}\, in the case of the MX$_1$ resonance. We then perform a Gaussian fit of the measured line shift as a function of the pump--probe delay in order to exctract the light shift amplitude at zero time delay. To determine the dependence of the light shift on the pump detuning, we perform for each pump wavelength a measurement at various pump powers as shown in Fig.\,\ref{fig4SI}\,{\bf b}. In this way, we can use relatively low power for a near resonant excitation and larger power at larger detunings, always making sure that we are in a regime of linear scaling with power.  In the figures of the main text, we plot the slope $\Delta/I_{\rm pump}$ computed from a linear fit at each pump detuning $\delta$. 

\section{Fit of the wavelength dependence}\label{App: Fit}
We discuss here our analysis of the wavelenght dependance of the light shift. Taking into account a single resonance, the light shift $\Delta$ can be expanded in a series of $1/\delta$ using perturbation theory \cite{combescot1992semiconductors}. The first order $1/\delta$ term comes from light-matter interaction, while exciton-exciton interactions contribute to higher order term, in $1/\delta^2$ or $1/(\delta_i\delta_j)$ when several modes contributes. In principle, a fine analysis of the detuning dependence of the light shift would enable to deconvolve the various contributions. In practice, this analysis can be challenging due to the finite range of detuning in which we can take reliable data. Indeed, for all resonances, we had to restrict the data to a window of $\approx[20,80]$\,meV. Going closer to resonance we face two issues: First, incoherent effects become more prominent as pump photons carry enough energy to generate a real population of excitons and hence are more likely to be absorbed. Second, the perturbative expansion of the light shift becomes inconsistent when $\Delta\sim\delta$. Conversely, going further away from resonance the signal reduces (given the available laser power) and the extraction of the light shift becomes unreliable. Furthermore, in that regime, the light--matter terms $\propto1/\delta$ becomes stronger compared to the interaction term $\propto1/\delta^2$ that we are interested in.

\subsection{Charge neutrality}
\begin{figure*}
    \centering
    \includegraphics[width=\textwidth]{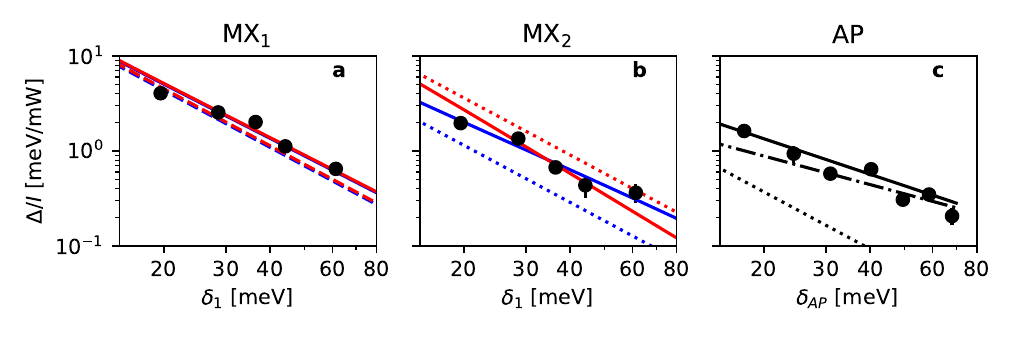}
    \caption{\textbf{Pump wavelength dependence.} From a fit we infer the origin of the light shift of MX$_1$ ({\bf a}), MX$_2$ exciton ({\bf b}) and of the attractive polaron at $\nu=1$ {\bf c}. 
    At charge neutrality ({\bf a} and {\bf b}), we are not able to deconvolve the contribution of the $u_{1,2}'$ and $k_{2}'$ fitting parameters (see Eq.\,[\ref{Eq: FitLightShiftMX2}]). We thus perform two fits, setting $k_{2}'$ to its upper and lower bounds, shown as solid blue and red line, respectively. For the MX$_1$ exciton ({\bf a}), the light shift is dominated by the  $2u_{1,1}n_1$ term (shown as dashed line) and barely sensitive to the MX$_2$-dependent terms, so both fit are nearly identical and we obtain a good estimate of $u_{1,1}$. For MX$_2$ ({\bf b}), both fits are in fair agreement with the data, but the contribution of the $2u_{1,2}n_1$ term (dotted line) is significantly different, and we are thus only able to provide a rather broad confidence interval for $u_{1,2}$. For the AP, a fit $A_\mathrm{AP }/\delta_1+B_\mathrm{AP}/\delta_\mathrm{AP}^2$  (solid black line, {\bf c}) suggests that the usual AC-Stark shift scaling as $A_\mathrm{AP }/\delta_1$ (dashed-dotted line) is occurring, instead of an interaction-driven shift scaling as $B_\mathrm{AP}/\delta_\mathrm{AP}^2$ (dotted line).} 
    \label{figSI_fit}
\end{figure*}
With the relatively small detuning we use (compared to the exciton Rydberg energy), the exciton-exciton interaction is expected to be the dominant contribution to the light shift \cite{uto2024}. This is indeed confirmed by our data which are not compatible with a $1/\delta$ law. Let us thus focus on the interaction induced light shift, which we generalize to our moiré system with two bright exciton modes. As shown in App.\,\ref{App: Same valley interactions}, the light shifts of MX$_1$ and MX$_2$ read
\begin{align}
    \Delta_1 &= 2u_{1,1}n_1+2u_{1,2}n_2+4k_{1}\sqrt{n_1n_2}\,,\\
    \Delta_2 &= 2u_{2,2}n_2+2u_{1,2}n_1+4k_{2}\sqrt{n_1n_2}\,.
\end{align}
with the densities scaling as $n_i\propto |\Omega_i\phi_i(0)|^2/\delta_i^2$. The ratio $r=|\Omega_1\phi_1(0)|^2/|\Omega_2\phi_2(0)|^2=\gamma_{\rm rad,1}/\gamma_{\rm rad,2}\approx0.48$ can be obtained from a fit of the reflection contrast (see App.\,\ref{App: TMM}). Using this independent estimate and performing a joint fit of $\Delta_{1,2}$ we reduce the number of fitting parameters to five, $A$, $u_{2,2}'$, $u_{1,2}'$, $k_{1}'$ and $k_{2}'$, where $u_{i,j}'=u_{ij}/u_{1,1}$, $k_i'=k_i/u_{1,1}$ and
\begin{align}
    \Delta_1 &= A\left(\frac{1}{\delta_1^2}+\frac{r^2u_{1,2}'}{\delta_2^2}+\frac{2rk_{1}'}{\delta_1\delta_2}\right)\,,\label{Eq: FitLightShiftMX1}\\
    \Delta_2 &=A\left( \frac{r^2u_{2,2}'}{\delta_2^2}+\frac{u_{1,2}'}{\delta_1^2}+\frac{2rk_{2}'^{2,2}}{\delta_1\delta_2}\right)\,.\label{Eq: FitLightShiftMX2}
\end{align}
Furthermore, within a Born approximation $u_{ij}^{kl}\propto\int dr\phi_i\phi_j\phi_k^*\phi_l^*$ (see App\,\ref{App: Same valley interactions}) and using the Cauchy-Schwarz inequality, we obtain the following bounds $0< u_{1,2}'<[u_{2,2}']^{1/2}$\,, $|k_{1}'|<[u_{1,2}]^{1/2}$\,, $|k_{2}'|<[u_{2,2}'u_{1,2}']^{1/2}$, which we enforce to improve the convergence of the fit.

For MX$_1$, see fig.\,\ref{figSI_fit}\,{\bf a}, the full fit reveals the dominant contribution to be the MX$_1$-MX$_1$ interaction. We obtain the fitting parameter $A$, which we can compare to the value obtain from fitting the light shift of a monolayer exciton (on a monolayer MoSe$_2$ region of the same device). We estimate $u_{1,1}/u_{\rm ex}\approx1.6\pm 0.2$, where $u_{\rm ex}$ is the monolayer exciton-exciton interaction (see App.\,\ref{App: InteractionStrength} for an absolute calibration). 

For MX$_2$, see fig.\,\ref{figSI_fit}\,{\bf b}, the fit is unable to reliably disentangle the contribution of the second and third terms of \ref{Eq: FitLightShiftMX2}, corresponding the sattering of an MX$_1$ exciton, without (1st term) or with (2nd term) a change of moiré band. Indeed, performing a fit with each of these two terms independently, we obtain in both case a reasonable agreement with the data over the full detuning range. We thus perform two fits, where the coupling $k_{2}'=\pm[u_{2,2}'u_{1,2}']^{1/2}$ staturates the Cauchy-Schwartz inequality. In this way we obtain the bounds $0.3\lessapprox u_{2,2}\lessapprox 0.8$.

\subsection{Moiré attractive polaron}
For the AP ({\bf c}), the light shift is typically smaller and more noisy due to the weakness of the transition. As a result, it is more difficult to discriminate a potential $1/\delta$ and $1/\delta^2$ dependence. Nevertheless, the fit does suggest that the former is here the leading contribution, consistent with the picture of an ensemble of distinguishable and non-interacting two-level systems, as argued in the main text.

\section{Transfer matrix simulation}\label{App: TMM}

\begin{figure}
    \centering
    \includegraphics[width=\columnwidth]{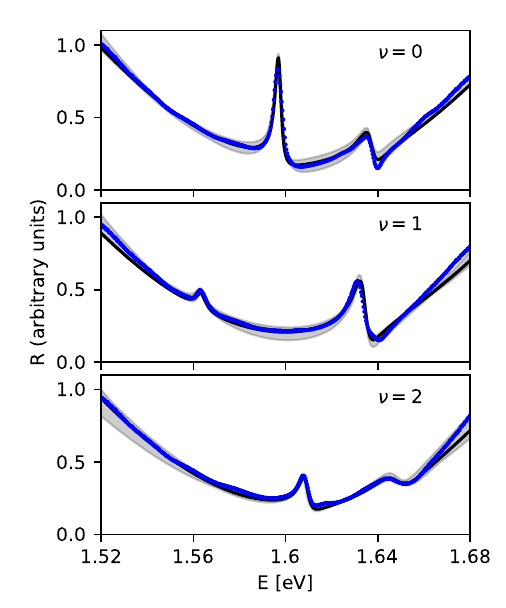}
    \begin{tabular}{ |c|c|c|c| } 
 \hline
 ~Resonance~ & ~$\nu$~ & ~$\hbar\gamma_{\rm rad}$ [meV]~ & ~$\hbar\gamma_\text{non-rad}$ [meV]~ \\ 
 \hline
 MX$_1$ & 0 & $1.0\pm0.2$ & $1.3\pm0.3$ \\ 
 MX$_2$ & 0 & $0.48\pm0.1$ & $2.8\pm0.4$\\ 
 AP & 1 & $0.27\pm0.1$ & $2.7\pm0.4$ \\ 
 MX$_2^\prime$ & 1 & $1.1\pm0.3$ & $2.9\pm0.4$\\ 
MX$_3$ & 2 & $0.52\pm0.1$ & $2.2\pm0.3$\\ 

 \hline    
\end{tabular}
    \caption{\textbf{Fit of the reflection spectrum using transfer matrix simulation.} The blue dots are the data, the black solid line is a fit (grey area span by varying the fitting parameters within the confidence interval).} 
    \label{figSI_TM}
    \begin{center}
\end{center}

\end{figure}

In order to infer the interaction strength of the various moiré excitons from their light shift, we need to estimate the exciton density that we generate, and hence the exciton oscillator strength. The latter can be obtained from a fit of the reflection spectrum. Such a fit needs to include the reflection of the electromagnetic field on the interfaces between the various dielectrics that make our van der Waals heterostructure. We do this using the transfer matrix method \cite{scuri2018large}. Two fitting parameters for the background are the hBN thickness $\approx31\,$nm and $\approx37\,$nm for the top and bottom layer and the hBN refractive index $n_{\rm hBN}\approx2.15$ \cite{laturia2018dielectric,kim2012synthesis}. Then, for each resonance, we have three additional fitting parameters, namely its energy, radiative and non-radiative decay rates. The results of this fit are shown in Fig.\,\ref{figSI_TM}. We show here the bare reflection spectrum, obtained using a light source that is to a very good approximation spectrally flat in the energy range shown in the figure. We are unable to obtain a perfect fit of the background, using the hBN thicknesses and refractive index as free parameters. The discrepancy that we observe, especially on the edge of the spectrum could be due to chromatic aberrations (although we are using a microscope apo\-chro\-matic objective to limit those). Nevertheless, in the center of the spectrum, we are able to reproduce our spectra very well at all fillings. 

\section{Interaction strength}\label{App: InteractionStrength}
\begin{figure}
    \centering
    \includegraphics[]{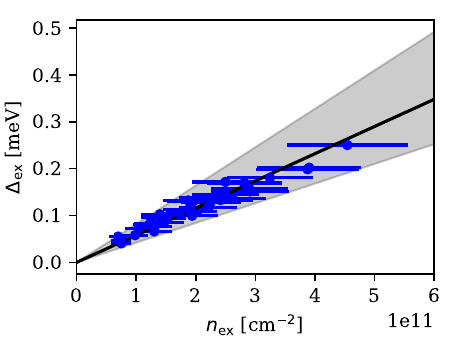}
    \caption{\textbf{Light shift as a function of the exciton density.} The blue dots are the data, obtained for various powers and detunings. The error bars come from the uncertainty of the exciton decay rates and on the power on the sample. The black solid line is a fit with the grey area span obtained by varying the fitting parameters within the confidence interval.} 
    \label{figSI_Interaction}
    \begin{center}
\end{center}

\end{figure}

The radiative ($\gamma_{\rm r}$) and non-radiative ($\gamma_{\rm nr}$) decay rates can be used to extract the density of excitons $n_{\rm ex}$ induced by the pump. Using the optical Bloch equations within the adiabatic approximation we have \cite{scuri2018large} 
\begin{align}
    n_{\rm ex}=\frac{2I_{\rm i}\gamma_{\rm r}}{(\gamma_{\rm r}+\gamma_{\rm nr})^2+\delta_{\rm ex}^2}\,,
\end{align}
where $I_{\rm i}$ is the photon flux. For reference, we first look at the exciton light shift $\Delta_{\rm ex}$ measured on a monolayer region of the sample, as a function of the exciton density, see Fig.\,\ref{figSI_Interaction}. We observe a linear dependence $\Delta_{\rm ex}=u_{\rm ex}n_{\rm ex}$, from which we extract the exciton--exciton interaction strength $u_{\rm ex}\approx \SI{0.06+-0.03}{\micro\electronvolt\micro\meter\squared}$.
The latter is compatible with previous measurements \cite{barachati2018interacting,scuri2018large,tan2020interacting,uto2024}, although significantly smaller than theoretical estimates,
$\sim 3\,E_\mathrm{ex} a_\mathrm{ex}^2 \sim \SI{1}{\micro\electronvolt\micro\meter\squared}$, where $E_\mathrm{ex}$ is the exciton binding energy and $a_\mathrm{ex}$ is the exciton Bohr radius \cite{ciuti1998interaction,shahnazaryan2017excitoninteraction}.

From the measurement of the light shift of the moiré excitons and polaron shown in Fig.\,\ref{figSI_fit} we obtain $u_{1,1}\approx 1.6\,u_{\rm ex}$ , $u_{1,2}\approx 0.7\,u_{\rm ex}$ and for the MX$_3$-MX$_3$ interaction, $u_{3,3}\approx 5.3\,u_{\rm ex}$. The enhancement of the interaction strength of MX$_1$ (charge neutrality) could stem from the partial confinement induced by the moiré potential. The larger enhancement for MX$_3$ reflects the polaronic nature of this resonance, as previously observed in a monolayer sample \cite{uto2024}.

\section{Light shift of MX$_3$ in the regime of large doping}\label{App: MX3}
\begin{figure*}
	\centering
	\includegraphics[width=\linewidth]{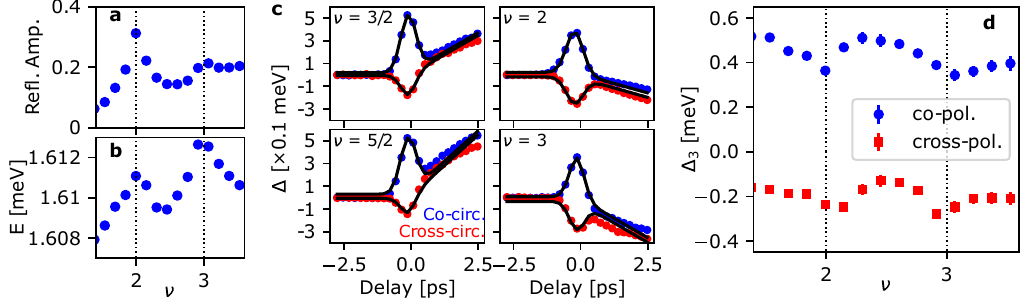}
        \caption{\textbf{Light shift of MX$_3$.} The resonance MX$_3$ shows distinctive kinks in its amplitude ({\bf a}) and energy ({\bf b}) at integer fillings of the moiré lattice with electrons, following the changes in the 2DES compressibility. The light shift of the moiré exciton MX$_3$ at various fillings ({\bf c}), is well fitted by a line on top of a Gaussian function (black line). The latter captures the coherent response of the system, which corresponds to an interaction-induced blue shift for co-circularly polarized pump and probe, and possibly to a red shift stemming from the coupling to the biexciton states in a cross-circular configuration. The amplitudes of these two shifts are sensitive to the polaron dressing of the exciton, and consequently are extremal at integer filling of the moiré lattice ({\bf d}). }
	\label{FigureMX3}
\end{figure*}

Even though MX$_3$ is the dominant excitonic resonance for electron filling factors $\nu \ge 1.5$, its identification has remained unclear. To gain insight, we investigate the nonlinear response of the  MX$_3$ resonance. Figure~\ref{FigureMX3}\,{\bf a},\,{\bf b} show the reflection amplitude and the energy of MX$_3$ for $1.5 \le \nu \le 3.4$ in the absence of a pump laser: Consistent with earlier observations, we find that the resonance energy as well as the reflection strength, or equivalently the oscillator strength, of MX$_3$ exhibit local maxima at integer fillings $\nu=2$ and $\nu = 3$. These features could be explained by partial suppression of dynamical dressing of excitons by electrons, when the two dimensional electron systems (2DES) is in an incompressible state\,\cite{Smolenski2019,Xu2020}. In contrast, when the electrons form a Fermi liquid ($\nu \neq 2,3$), the dynamical dressing of MX$_3$ is more effective and results in a red shift together with a reduction of the oscillator strength. 

Figure~\ref{FigureMX3}\,{\bf c} shows the light shift as a function of pump--probe delay $\tau$ for four representative filling factors obtained for $\delta_3 = 80$~meV. We observe that the the light shift for $\tau \simeq 0$ indicates repulsive (attractive) interactions between same (opposite) valley MX$_3$ excitons generated by co- (cross-) circularly polarized pump--probe fields. While attractive interactions between opposite valley excitons has been reported before, it is surprising that the magnitude of the light shift is comparable in the two cases. The attractive interactions for the cross-polarized configuration may be explained through a near-resonant two-photon (pump+probe) excitation of the biexciton resonance at $\omega_\mathrm{XX}$. Verification of this hypothesis could be achieved by changing the pump detuning $\delta_3$ so as to probe both $\delta_3 \le \omega_\mathrm{XX}$ and  $\delta_3 \ge \omega_\mathrm{XX}$, since for the latter case, the biexciton-mediated interactions would become repulsive. As we had to choose $\delta_3 < \omega_\mathrm{XX}$ to avoid strong background absorption, we could not verify the role of biexciton in the measured light shift.

In contrast to the light shift measurements in the charge-neutFral regime, we find that the pump pulse results in a MX$_3$ line shift that increases linearly with $\tau$ for $0.2 \le t \le 3.0$\,ps. Moreover, the linear shift at a given $\nu$ is identical for co- and cross-circularly polarized pump--probe configurations, but has a different sign for compressible ($\nu \neq 2,3$) and incompressible ($\nu=2,3$) electron states. We tentatively explain this feature by generation of free carriers by non-resonant absorption of pump-photons that change the electron density for timescales well exceeding the pump duration. 

Since the MX$_3$ resonance energy has maxima (minima) for $\nu = 2,3$ ($\nu = 3/2,5/2$), any pump-induced change in electron density will result in a red (blue) shift of the resonance energy. While we do not understand why the red (blue) shift depends linearly on $\tau$ for $\tau > \tau_\mathrm{pump}$, we speculate that pump-induced charges are initially generated in high-energy bands and that they influence the nonlinear response only as they relax to the lowest-energy available moiré band.

We also observe in Fig.\,\ref{FigureMX3}\,{\bf c} that the magnitude of the light shift for the co-circularly polarized pump--probe configuration is smaller for incompressible states.  Plotting $\Delta_3$ as a function of $\nu$ (Fig.\,\ref{FigureMX3}\,{\bf d}) shows that the blue shift is indeed minimal for $\nu = 2,3$. This is at first glance surprising given that the oscillator strength, and consequently the generated MX$_3$ population, is maximal for these incompressible states. However, it was recently shown that interactions between attractive exciton-polarons\,\cite{uto2024,muir2022interactions,tan2020interacting} mediated by their dressing cloud, are dramatically enhanced compared to those of bare excitons. Such an enhancement of interaction strength may overcome the reduction of the oscillator strength of MX$_3$ for $\nu \neq 2,3$. This tentative explanation suggests that the MX$_3$ mode may be identified as a second attractive polaron mode where the exciton is dressed by electrons in the upper moiré band. Last but not least, we find that the red shift of MX$_3$ in the cross-circularly polarized configuration is maximal when the electronic state is incompressible; we currently do not have an explanation for this observation.

\section{Light shift of MX$^\prime_2$ at $\nu = 1$}

\begin{figure}
    \centering
    \includegraphics[width=\linewidth]{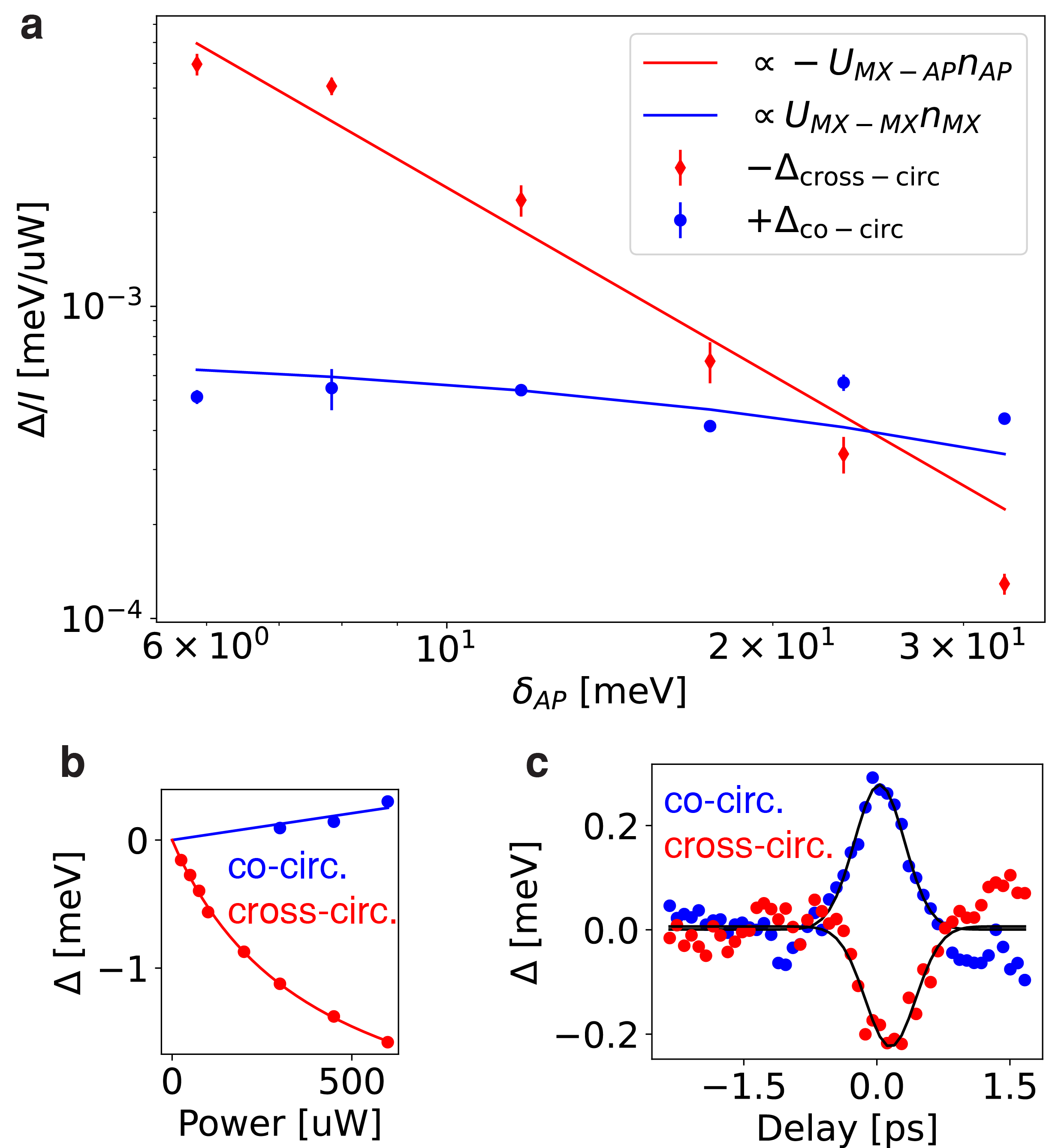}
    \caption{\textbf{Light shift of MX$_2^\prime$ at $\nu = 1$.} (\textbf{a}) Detuning dependence of the light shift of the resonance MX$^\prime_2$ at a filling $\nu = 1$. Red dots correspond to cross-polarized pump and probe and show a red shift of the MX$^\prime_2$ resonance which can be well fitted (red line) by a dependence $\propto1/\delta_{\rm AP}^2$ stemming from MX$^\prime_2$--AP interactions. Blue dots correspond to co-polarized pump and probe and show a blue shift of the MX$^\prime_2$ resonance which can be well fitted (blue line) assuming MX$_2^\prime$--MX$^\prime_2$ interactions and a scaling  $\propto1/\delta_{\rm MX_2^\prime}^2$. (\textbf{b}) Lightshift of MX$_2^\prime$ as a function of pump power in different polarizations. Cross-circular polarization shows a distinct red shift; the deviation from a linear dependence is due to saturation of the AP population at large powers. Co-linear polarization shows a smaller blue shift which is well in the linear regime. (\textbf{c}) MX$_2^\prime$ shift vs. pump--probe time delay in co-circular (blue) for a pump power of \SI{500}{\micro\watt} and cross-circular (red) polarizations for \SI{50}{\micro\watt}.}
    \label{fig3SI}
\end{figure}

At a unity filling $\nu=1$ of the moiré potential, we observe two bright resonances, the AP which we discussed in detail in the main text, and MX$_1^\prime$, emerging from MX$_2$, and which we now investigate.
Depending on the pump and probe polarizations, we observed different behaviors. In co-circular polarization, we obtain the usual blue shift which we attribute to MX$_2^\prime$--MX$_2^\prime$  interactions. Contrary to other resonances, we cannot confirm this claim by an analysis of the detuning dependence. Indeed, we observed strong incoherent behavior for a pump blue detuned from the AP, and we therefore only explored the red detuned situation. Specifically, we explore the range $\delta_{\rm MX2'}\approx [80,110]$\,meV in which we observe no significant evolution of the light shift, as expected from a scaling as $1/\delta_{\rm MX_2^\prime}^2$, see Fig.~\ref{fig3SI}\,\textbf{a}. By contrast, in cross-circular polarization we observe a distinct redshift which diverges close to the AP resonance, in excellent agreement with a $1/\delta_{\rm AP}^2$ scaling and suggesting an attractive interaction between AP and MX$_2^\prime$ in opposite valleys. We point out that a similar behavior was observed in a monolayer system, and tentatively attributed to the reduction of the phase space filling upon the generation of an AP, for an opposite valley exciton \cite{uto2024}. The pump power dependence (\textbf{b}) of the light shift shows an interesting behavior for a near-resonant excitation of the AP $\delta_{\rm AP}\approx8$\,meV. In that case, at high intensity, we expect a saturation of the AP density as the moiré potential is filled up, as described in the main text (although here, keeping a finite $\delta_\mathrm{AP}$ we are unable to fully saturate the transition). Indeed, we observe a sub-linear increase of the cross-polarized light shift attributed to AP--MX$_2^\prime$ interactions. On the contrary, the co-polarized light shift which we attribute to MX$_2^\prime$--MX$_2^\prime$ interactions is linear in the pump power, as the density of MX$_2^\prime$ remains far from saturation of the moiré lattice ($\delta_{{\rm MX}_2^\prime}\approx80\,$meV).

\section{Theoretical model} \label{App:TheoryModel}
In this section, we primarily provide a detailed description of the theoretical model used to analyze the charge-neutral, cross-circular polarization data. The foremost goal of this model is to qualitatively capture the energy shifts observed in the MX$_1$ and MX$_2$ exciton resonances when subjected to the influence of a pump laser. To this end, we calculate the exciton-exciton $T$ matrix in the presence of a moiré potential, which is inspired by a recent treatment of two interacting atoms in a 2D square optical lattice~\cite{Adlong2024}. This approach allows us to capture the essential physics of the interaction under the influence of the pump.

We conclude the section with a brief discussion on a model for the co-circular polarization data. For ease of notation we set $\hbar= 1$.

\subsection{Model for interactions of two distinguishable excitons}
We begin by modeling the interactions between two distinguishable rigid (1s) excitons in the presence of moiré. The Hamiltonian we consider consists of four terms
\begin{align} \label{Eq:2bodyHam}
	H = H_{X \uparrow} + H_{X \downarrow} + H_d+ V.
\end{align}
The first two terms represent the exciton Hamiltonians for each valley ($\sigma = { \uparrow, \downarrow }$):
\begin{align} \label{Eq:model}
	H_{X \sigma} = \sum_{\Kv} \epsilon_{\Kv,X} \Xhd_{\Kv \sigma} \Xh_{\Kv \sigma}+ \sum_{\Kv \Qv} \tilde V_X(\Qv) \Xhd_{\Kv + \Qv \sigma} \Xh_{\Kv \sigma}.
\end{align}
Here, $\Xh_{\Kv \sigma}$ annihilates an exciton of type $\sigma$ with energy $\epsilon_{\Kv,X} = \epsilon_X + |\Kv|^2/2m_X$, where $m_X$ is the exciton mass and $\epsilon_X$ is the 1s exciton energy. The term $\tilde V_X$ is the Fourier transform of the exciton moiré potential, which is approximated in real space by\,\cite{wu2017topological}
\begin{align}
V(\vb{x}) = \sum_{j=1}^6 V_j e^{i \bv_j \cdot \vb{x}},
\end{align}
where $\bv_j$ are the first six reciprocal lattice vectors. The three-fold symmetry and the realness of the periodic potential imply that $V_1 = V_3 = V_5$, $V_2 = V_4 = V_6$, and $V_1 = V_4^*$, parametrized by $V_1 = V e^{i \psi}$, with $V$ determining the potential depth and $\psi$ its shape.

Table~\ref{tab:particle_parameters} summarises the relevant parameters for our system as determined by large-scale Density Functional Theory (DFT) calculations\,\cite{Ciorciaro2023}. Here, we also include the exciton moiré parameters which are determined by assuming a tightly bound electron-hole pair such that $V_X e^{i \psi_X} \equiv V_e e^{i \psi_e} + V_h e^{i \psi_h}$, where $V_i$ and $\psi_i$ are the moiré parameters for the exciton ($i=X$), electron ($i=e$) and hole ($i=h$).
\begin{table}
\centering
\begin{tabular}{>{\arraybackslash}m{2cm}ccc}
\textbf{Particle} & \textbf{$V$ [$\rm meV$]} & \textbf{$\psi$} & \textbf{$m$ [$m_e$]} \\
\hline
Electron & $-6.3$ & $0^{\circ}$ & $0.45$ \\
Hole & $1.9$ & $59^{\circ}$ & $0.55$ \\
Exciton & $5.6$ & $163^{\circ}$ & $1$ \\
\hline
\end{tabular}
\caption{Parameters for electron, hole, and exciton from density functional theory calculations~\cite{Ciorciaro2023}. The mass of the particles is given in units of the bare electron mass $m_e$. The assumed moiré length is $a_M = 8.2$ nm.}
\label{tab:particle_parameters}
\end{table}

The third term in the Hamiltonian describes the closed-channel molecule, which mediates interactions between excitons:
\begin{align} \label{Eq:InteractionHam}
	H_d = \sum_{\Kv} (\epsilon_{\Kv,d}+\delta_{\rm cc}) \dhd_{\Kv} \dh_{\Kv} + \sum_{\Kv \Qv}2 \tilde V_X (\Qv) \dhd_{\Kv+\Qv} \dh_{\Kv},
\end{align}
where $\dh_{\Kv}$ annihilates a closed-channel molecule with energy $\epsilon_{\Kv,d} = 2 \epsilon_X + |\Kv|^2/2M$ ($M = 2m_X$) and detuning $\delta_{\rm cc}$. The closed-channel experiences both exciton potentials, hence the factor of two in the moiré potential.

The interactions are mediated according to
\begin{align} \label{Eq:ActualInteractionHam}
	V = \frac{g}{\sqrt{\area}} \sum_{\Kv \Qv} \chi(\Kv)( \dhd_{\Qv} \Xh_{\Qv-\Kv, \uparrow}  \Xh_{\Kv, \downarrow} + \rm{h.c.}),
\end{align}
where $\chi(\Kv)$ regularizes the ultraviolet (UV) divergence and $\area$ is the system area. Throughout this section we will use $\chi(\Kv) = \Theta(\Lambda - |\Kv|)$, where $\Lambda$ is the UV cutoff. The bare coupling $g$ and detuning $\delta_{\rm cc}$ are renormalized as~\cite{LevinsenStronglyInteracting}
\begin{align}
	\frac{\delta_{\rm cc}}{g^2} =  \frac{1}{\area} \sum_{\Kv} \chi(\Kv) \frac{1}{E_{\rm BX}+2 \epsilon_{\Kv,X} },
\end{align}
with $E_{\rm BX}$ being the energy of the biexciton \textit{without moiré}.

We point out that the two-channel model is equivalent to contact interactions with coupling constant $U = - g^2 / \delta_{\rm cc}$ in the single-channel model limit ($\delta_{\rm cc},g\to \infty$). Throughout this work, we exclusively work in the single-channel model limit since, for our purposes, the two-channel model is only used as a tool to simplify the calculation of the exciton-exciton $T$ matrix.

\subsection{T matrix}

To study the interactions we calculate the $T$ matrix, which provides an exact solution to the full two-body problem. To begin we introduce the free exciton and closed-channel Green's functions:
\begin{align}
		\hat G^{(0)}(E) &= \frac{1}{E-\hat H_{X \uparrow} - \hat H _{X, \downarrow}}\\
	\hat D^{(0)}(E) &= \frac{1}{E-\hat H_d}.
\end{align}
The two-body $T$ matrix is given by the infinite series
\begin{align}
	T &= \hat V \Dfreeh \hat V + \hat V \Dfreeh \hat V \Gfreeh \hat V \Dfreeh \hat V + \dots \nonumber \\
	&= \hat V \left( \Dfreeh  + \Dfreeh \hat V \Gfreeh \hat V \Dfreeh + \dots  \right) \hat V \nonumber \\
	&= \hat V \Dh \hat V
\end{align}
where we have suppressed the energy dependencies for brevity and $\hat{D}(E)$ is the closed-channel Green's function. Thus, by finding the closed-channel Green's function, we can immediately calculate the $T$ matrix. This approach simplifies the calculation and provides easier access to the biexciton energies, which are both the poles of the $T$ matrix and $\hat{D}(E)$.

The closed-channel Green's function is given by
\begin{subequations}\label{eq:closedchG}
	\begin{align}
		\Dh(E) 
		&=\frac{1}{[\Dfreeh (E)]^{-1} - \frac{g^2}{\area} \Pih (E)},
	\end{align}
\end{subequations}
where we have introduce the polarization bubble, which has matrix elements
\begin{align} \label{Eq:PiElements}
	\Pi^{\qv}_{\lambda \lambda'} =&\sum_{\kv,\lambda_\uparrow,\lambda_\downarrow}  \mathcal{V}^{\qv,\lambda}_{\kv \lambda_\uparrow \lambda_\downarrow} \frac{1}{E - E_{\kv, \qv; \lambda_\uparrow, \lambda_\downarrow}}
	\mathcal{V}^{\qv,\lambda' *}_{\kv \lambda_\uparrow \lambda_\downarrow},
\end{align}
where 
$E_{\kv, \qv; \lambda_\uparrow, \lambda_\downarrow} \equiv E^{\uparrow}_{\qv-\kv,\lambda_\uparrow} + E^{\downarrow}_{\kv,\lambda_\downarrow} $ and
\begin{align} \label{Eq:MatElements}
	&\mathcal{V}^{\qv \lambda_d}_{\kv \lambda_\uparrow \lambda_\downarrow}   \equiv \bra{\qv, \lambda_d, d} V \ket{\qv-\kv, \lambda_\uparrow, \uparrow; \kv,\lambda_\downarrow, \downarrow} \sqrt{\area} /g \nonumber 
\end{align}
Here we have introduced a Bloch basis, which satisfy
\begin{align}
H_{X, \sigma} \ket{\kv, \lambda, \sigma} &= E^{\sigma}_{\kv, \lambda} \ket{\kv, \lambda, \sigma} \\
H_d \ket{\kv, \lambda, d} &= (E^d_{\kv, \lambda} + \delta_{\rm cc}) \ket{\kv, \lambda, d},
\end{align}
where $\kv$ is the quasi-momentum and $\lambda$ is the band index.

\subsection{Modeling lights shifts}
Within our model for cross-circular polarization data, we consider the interaction induced shift of the probe exciton (denoted by $\uparrow$) by virtual excitons ($\downarrow$) created by the pump. To incorporate the effects of the pump laser, we consider the following modification to the Hamiltonian $\hat H_{X \downarrow}$\,\cite{combescot1992semiconductors}
\begin{align} 
	H_{X \downarrow} &= \sum_{\kv, \lambda} E^\downarrow_{\kv, \lambda} \Xhd_{\kv,\lambda,\downarrow} \Xh_{\kv,\lambda,\downarrow} + \Omega{}^* e^{-i \omega_L t} \sum_{\lambda} \phi_{\zerov}^{(\zerov, \lambda, \downarrow)*} \Xhd_{\zerov, \lambda, \downarrow} \nonumber \\
 &{}\qquad + \Omega e^{i \omega_L t}\sum_{\lambda} \phi_{\zerov}^{(\zerov, \lambda, \downarrow)} \Xh_{\zerov, \lambda, \downarrow},
\end{align}
where $\phi^{(\zerov, \lambda, \downarrow)}_{\zerov} \equiv \vacbra \Xh_{\zerov, \downarrow} \ket{\zerov, \lambda, \downarrow}$ and $\Xh_{\kv,\lambda, \downarrow}$ annihilates a $\downarrow$ exciton with quasi-momentum $\kv$ and band index $\lambda$. Here, $\Omega$ represents the light-matter interaction strength; we assume that the light couples only to the zero-momentum exciton and use the rotating wave approximation ($|\epsilon_X - \omega_L|\ll \epsilon_X + \omega_L$). By moving into the rotating frame, it can be seen that the excitons form a coherent state, $\ket{\Phi} = \hat{\underline{D}}(\beta) \ket{0}$ (ignoring normalization), where we have introduced the multi-mode shift operator
\begin{align}
\hat{\underline{D}}(\beta) = \exp[ \sum_{\lambda} \left( \beta_\lambda \Xhd_{\zerov, \lambda, \downarrow} - \beta_\lambda^* \Xh_{\zerov, \lambda, \downarrow}\right)],
\end{align}
and  $\beta_\lambda = -\phi_{\zerov}^{(\zerov, \lambda, \downarrow)*} \Omega^*/(E^\downarrow_{\zerov, \lambda} - \omega_L)$

We can then approximate this energy shift of the $\lambda$-th resonance of the $\uparrow$ probe exciton using the $T$ matrix:
\begin{align} \label{Eq:ExcitonShift}
    \Delta_\lambda (\omega_L) \simeq  \sum_{\lambda_\downarrow \lambda_\downarrow'} \beta_{\lambda_\downarrow} T^{\lambda \lambda_\downarrow; \zerov}_{\lambda \lambda'_\downarrow; \zerov}(E_{0,\lambda} + \omega_L + 2 i \gamma)\beta^*_{\lambda_\downarrow'},
\end{align}
with $\gamma \approx 1$ meV the inverse lifetime of the exciton. Here, we have the matrix elements of the $T$ matrix defined by
\begin{align} \label{Eq:TmatrixElements}
    T^{\lambda_\uparrow \lambda_\downarrow; \zerov}_{\lambda_\uparrow' \lambda'_\downarrow; \zerov}(E) \equiv \bra{0} \Xh_{\zerov \lambda_\downarrow \downarrow} \Xh_{\zerov \lambda_\uparrow \uparrow} \hat T(E) \Xhd_{\zerov \lambda_\uparrow' \uparrow} \Xhd_{\zerov \lambda_\downarrow' \downarrow} \ket{0},
\end{align}
In Eq.~\eqref{Eq:ExcitonShift} we explicitly include the laser frequency $\omega_L$ as a parameter in the self energy to emphasize its significance: we observe that the laser light shifts the scattering off-shell, similar to the effect discussed in the context of polariton-electron scattering~\cite{Kumar2023}. It is important to note that Eq.~\eqref{Eq:ExcitonShift} captures the ac Stark effect, whereby the shift can change sign when the laser frequency $\omega_L$ is tuned into resonance with the biexciton bound state, i.e., $E_{\zerov, \lambda} = \omega_L - E_{\rm{BX}; \alpha}$, where $E_{\rm{BX}; \alpha}$ denotes the energy of a zero quasi-momentum biexciton state. However, unlike the simpler case of a monolayer TMD, this sign change does not occur at all biexciton energies, as not all biexciton states will couple to the $\lambda^{\rm th}$  exciton mode. We can therefore expect that in the experiments, which only observe shifts in the optically active MX$_1$ and MX$_2$ modes, it will not be possible to detect all biexciton states.

The magnitude of the shift in Eq.~\eqref{Eq:ExcitonShift} depends on our choice of $\Omega/\sqrt{\area}$. Since our primary interest is in the zero crossings of the shift (and not its absolute value) we use a simple estimate for this quantity. In particular, we take $\Omega/\sqrt{\area} = \sqrt{2}$ meV/nm, which is chosen to correspond to a reasonable induced exciton density of $5 \times 10^{11} \, \rm{cm}^{-2}$ of 1s excitons at a detuning of $20$ meV (in the absence of moiré).

\subsection{Same valley interactions}\label{App: Same valley interactions}
To conclude our work we briefly consider interactions from excitons in the same valley, which do not support a bound state. Owing to this, we simply take the Born approximation of the $T$ matrix assuming repulsive contact interactions. Focusing on zero quasi-momentum, the interactions between the excitons are given by
\begin{align} \label{Eq:ContactIntSameValley}
    \hat{\mathscr{V}} = \frac{u_{\rm ex}}{2} \sum U^{\lambda_1 \lambda_2}_{\lambda_3 \lambda_4} \Xhd_{\zerov, \lambda_1, \downarrow}\Xhd_{\zerov, \lambda_2, \downarrow} \Xh_{\zerov, \lambda_3, \downarrow} \Xh_{\zerov, \lambda_4, \downarrow}.
\end{align}
Here $u_{\rm ex}$ is the strength of the repulsion of two $\Kv =0$ excitons (i.e., in the absence of moiré), and we have introduced the overlap integral
\begin{align}
    U^{\lambda_1 \lambda_2}_{\lambda_3 \lambda_4} = \area \int d^2 r \, \varphi_{\lambda_1,\downarrow}^*  (\rv) \varphi_{\lambda_2,\downarrow}^* (\rv) \varphi_{\lambda_3,\downarrow} (\rv) \varphi_{\lambda_4,\downarrow} (\rv)
\end{align}
where $\varphi_{\lambda,\downarrow}$ is the Bloch wavefunction in \textit{real space} at zero quasi-momentum with band index $\lambda$ and valley index $\downarrow$. In order to derive the mean-field Hamiltonian given in the main text, we restrict ourselves to the optically bright excitonic states MX${}_1$ ($\lambda=0$) and MX${}_2$ ($\lambda=3$)
We furthermore use the fact that $\beta_\lambda / \sqrt{\area} = \kappa_\lambda \sqrt{n_{\lambda}}$ (with $\kappa_\lambda = e^{i \rm{arg}(- \phi_{\zerov}^{(\zerov,\lambda, \downarrow)}{}^* \Omega^*)}$) and $\bra{\Phi} \Xhd_{\zerov, \lambda, \downarrow} \Xh_{\zerov, \lambda', \downarrow} \ket{\Phi} = \beta_{\lambda}^* \beta_{\lambda'}$. Decomposing the interaction in Eq.~\eqref{Eq:ContactIntSameValley} using mean-field Hartree-Fock theory then yields Eq.~\eqref{Eq:MFShifts}. We point out that for contact interactions with bosons, the Hartree and Fock terms are identical in magnitude and sign. The relevant overlap integrals are
\begin{subequations}
    \begin{align}
    u_{1,1} /u_{\rm ex} &= U^{0,0}_{0,0} = 2.2\\
    u_{2,2} / u_{\rm ex} &= U^{3,3}_{3,3}  = 1.9 \\
    u_{1,2} / u_{\rm ex} &= U^{0,3}_{0,3} = 1.3\\
    k_{1}/ u_{\rm ex} &= \kappa_0^3 \kappa_3 U^{0,0}_{0,3}  = -1.2\\
    k_{2} / u_{\rm ex}&= \kappa_0 \kappa_3^3 U^{3,3}_{3,0}  = -0.4,
\end{align}
\end{subequations}
where we have assumed that $\kappa$ are real. All other interaction terms can be derived from these five, since the overlap integral only depends only on the number of each index (e.g., $U^{1,1}_{1,0} = U^{1,1}_{0,1} = U^{1,0}_{1,1} = U^{0,1}_{1,1}$).

\bibliographystyle{apsrev4-2}

%

\end{document}